\documentclass[a4paper,10pt,twoside]{cpc-hepnp}

\usepackage{multicol}
\usepackage{graphicx}
\usepackage{booktabs}
\usepackage{amssymb,bm,mathrsfs,bbm,amscd}
\usepackage[tbtags]{amsmath}
\usepackage{lastpage}

\begin{document}

\title{Study of spin sum rules (and the strong coupling 
constant at large distances)\thanks{Supported by the U.S. 
Department of Energy (DOE). The Jefferson Science Associates (JSA) 
operates the Thomas Jefferson National Accelerator Facility for the DOE 
under contract DE-AC05-84ER40150. }}

\author{%
      A. Deur$^{1)}$\email{deurpam@jlab.org}%
}
\maketitle

\address{%
1~(Thomas Jefferson National Accelerator Facility, 
Newport News, VA 23606\\
}

\begin{abstract}
We present recent results from Jefferson Lab on sum 
rules related to the spin structure of the nucleon. We 
then discuss how the Bjorken sum rule with its connection to the
Gerasimov-Drell-Hearn sum, allows us to conveniently define an 
effective coupling for the strong force at all distances.
\end{abstract}



\begin{multicols}{2}

\section{Introduction}
The information on the longitudinal spin structure of the nucleon is contained in the
$g_1(x,Q^2)$ and $g_2(x,Q^2)$ spin structure functions, with $Q^2$ the squared 
four-momentum transfered from the beam to the target, and $x=Q^2/(2M \nu)$ the
Bjorken scaling variable ($\nu$ is the energy transfer and $M$ the nucleon mass). 
The variable $Q^2$ indicates the space-time scale at which the nucleon is probed and
$x$ is interpreted in the parton model as the fraction of nucleon momentum carried by 
the struck quark.
 
Although  spin structure functions are the basic observables for nucleon spin studies, 
considering their integrals taken over $x$ is advantageous because of  
resulting simplifications. More importantly, such integrals are at the core
of the relation dispersion formalism. Relation dispersions relate the integral over
the imaginary part of a quantity to its real part. Expressing the 
imaginary part in function of the real part using the optical theorem yields 
\emph{sum rules}. When additional hypotheses are used, such as a low energy theorem
or the validity of Operator Product Expansion (OPE), the sum rules relate the integral 
to a static property of the target. If the static property is well known, the 
verification of the sum rule provides a check of the theory and hypotheses used in the 
sum rule derivation. When the property is not known because e.g. it is difficult to 
measure directly, sum rules can be used to access them. In that case, the theoretical
framework used to derived the sum rule is assumed to be valid. Details on integrals 
of spin structure functions and sum rules are given e.g. in the review~\cite{review moments}. 

Several spin sum rules exists. We will focus on the Bjorken sum rule~\cite{Bjorken} and the 
Gerasimov-Drell-Hearn (GDH) sum rule~\cite{GDH}. In this paper, we will 
consider the $n$-th Cornwall-Norton moments: $\int_{0}^{1}dx g_{1}^{N}(x,Q^2) x^n$, with
$N$ standing for proton or neutron, and 
write the first moments as $\Gamma_{1}^{N}(Q^2)\equiv\int_{0}^{1}dx g_{1}^{N}(x,Q^2)$.

\section{The generalized Bjorken and GDH sum rules}
The Bjorken sum rule~\cite{Bjorken} relates the integral over $(g_1^p-g_1^n)$ 
to the nucleon axial charge $g_A$. This relation has been essential for understanding the 
nucleon spin structure and establishing, \emph{via} its 
$Q^2$-dependence, that Quantum Chromodynamics (QCD) describes 
the strong force when spin is included.
The Bjorken integral has been measured in polarized deep inelastic
lepton scattering (DIS) at SLAC, CERN and DESY and at moderate $Q^2$ at Jefferson Lab 
(JLab), see Refs.~\cite{e142} to ~\cite{RSS}. 
In the perturbative QCD (pQCD) domain (high 
$Q^2$) the sum rule reads:
\begin{eqnarray}
\label{eq:bj(Q2)}
\Gamma_{1}^{p-n}(Q^2)\equiv\int_{0}^{1}dx
\left( g_{1}^{p}(x,Q^2)-g_{1}^{n}(x,Q^2) \right)=
\hspace{1cm}\\
\frac{g_{A}}{6}\left[1-\frac{\alpha_{s}}{\pi}-3.58
\frac{\alpha_{s}^{2}}{\pi^{2}}-
20.21\frac{\alpha_{s}^{3}}{\pi^{3}}+...\right]+
\sum_{i=2}^{\infty}{\frac{\mu_{2i}^{p-n}(Q^{2})}{Q^{2i-2}}} \nonumber
\end{eqnarray}
where  $\alpha_s(Q^2)$ is the strong coupling strength. The bracket term (known 
as the leading twist term) is mildly dependent on $Q^2$ due
to pQCD soft gluon radiation. The other term contains non-perturbative 
power corrections (higher twists). These are quark and gluon 
correlations describing the nucleon structure
away from the large $Q^2$ (small distances) limit.

The generalized Bjorken sum rule has been derived for small distances. 
For large distances, in the $Q^2 \rightarrow 0$ limit, one finds the 
generalized GDH sum rule. The sum rule was first derived at $Q^2=0$:

\begin{equation}
\int_{\nu_{0}}^{\infty}\frac{\sigma_{1/2}(\nu)-\sigma_{3/2}(\nu)}
{\nu}d\nu=-\frac{2\pi^{2}\alpha\kappa^{2}}{M_t^{2}}\label{eq:gdh}
\end{equation}

where $\nu_{0}$ is the pion photoproduction threshold, $\sigma_{1/2}$ and 
$\sigma_{3/2}$ are the helicity dependent photoproduction cross sections 
for total photon plus target helicities 1/2 and 3/2, $\kappa$ is the anomalous
magnetic moment of the target while $S$ is its spin and $M_t$ its mass. $\alpha$
is the fine structure constant.

Replacing the photoproduction cross sections by the electroproduction ones 
generalized the left hand side of Eq.~\ref{eq:gdh} to any $Q^2$. Such generalization 
depends
on the choice of convention for the virtual photon flux, see e.g. 
ref.~\cite{review moments}. X. Ji and J. Osborne~\cite{ji01} showed that the 
sum \emph{rule} itself  (i.e. the whole Eq.~\ref{eq:gdh}) can be generalized as:

\begin{equation}
\frac{8}{Q^2}\int_0^{x^-} g_1dx=s_1(0,Q^2)\label{eq:gdh*}
\end{equation} 

where $S_1(\nu,Q^2)$ is the spin dependent Compton amplitude. This generalization
of the GDH sum rule makes the connection between the Bjorken and GDH generalized sum 
rules evident: GDH$=\frac{Q^2}{8} \times $Bjorken. 

The connection between the GDH and Bjorken sum rules allows us in principle to 
compute the moment $\Gamma_1$ at any $Q^2$. Thus, it provides us with a choice
observable to understand the transition of the strong force  from small to large distances.

\section{Experimental measurements of the first moments}

Results from experimental measurements from SLAC~\cite{SLAC}, CERN~\cite{SMC}, 
DESY~\cite{HERMES} and JLab~\cite{eg1a proton}$^-$\cite{RSS} of the first moments are shown in 
Figure~\ref{gammas}.

\end{multicols}
\ruleup
\begin{center}
\includegraphics[width=17cm]{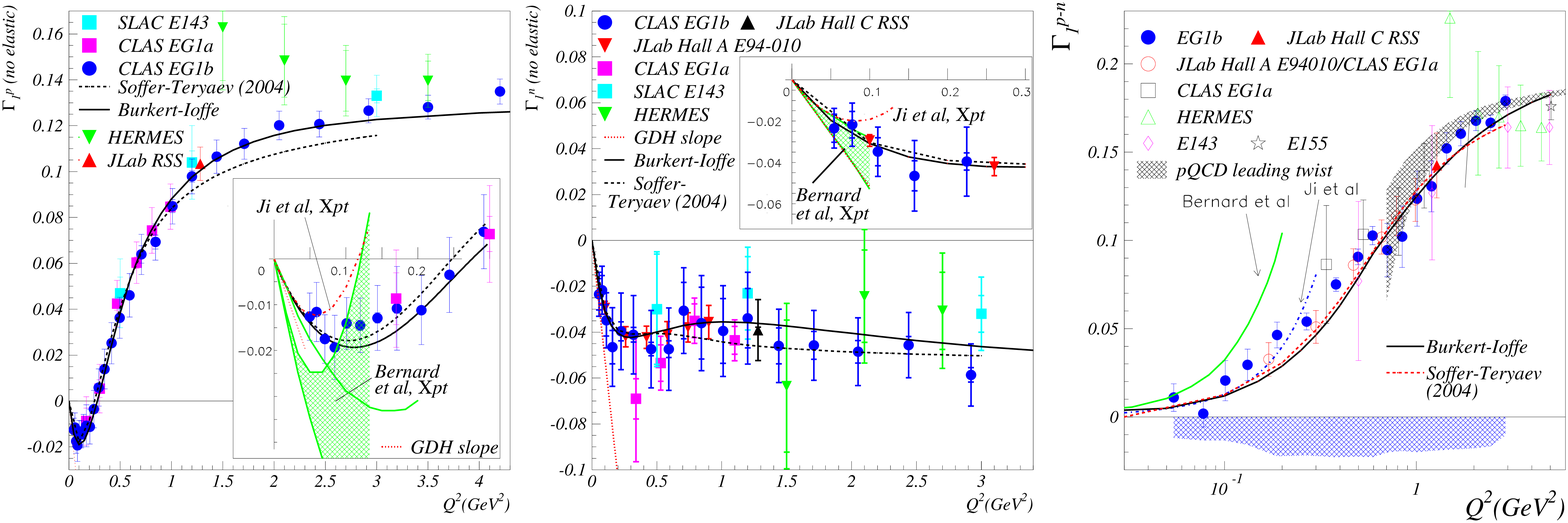}
\figcaption{\label{gammas} 
(Color online) Experimental data from SLAC, CERN, DESY and JLab at low and intermediate 
$Q^2$ on $\Gamma_1^p$ (left), $\Gamma_1^n$ (center) and $\Gamma_1^{p-n}$ (right).}
\end{center}
\ruledown

\begin{multicols}{2}

There is an excellent mapping of the moments at intermediate $Q^2$ and enough data points
a low $Q^2$ to start testing the Chiral Perturbation Theory ($\chi PT$), the effective 
theory strong force at large distances. In particular, the Bjorken sum is 
important for such test because the (p-n) subtraction cancels the $\Delta_{1232}$ 
resonance contribution which should make the $\chi PT$ calculations significantly more 
reliable~\cite{Burkert Delta}.
The comparison between the data at low $Q^2$ and $\chi PT$ 
calculations~\cite{meissner}$^,$\cite{Ji chipt}  can be seen 
more easily in the insert in each plot of Fig.~\ref{gammas}. The 
calculations assume the $\Gamma_1$ slope at $Q^2$=0 from the GDH sum rule prediction. 
Consequently, $\chi PT$ calculates the deviation from the slope and this is what one 
should test. A meaningful comparison is provided by fitting the lowest data points
using the form $\Gamma_1^N=\frac{\kappa_N^2}{8M^2}Q^2+aQ^4+bQ^6...$ and compare the
obtained value of $a$ to the values calculated from $\chi PT$. Such comparison has been 
carried out for the proton, deuteron~\cite{EG1b moments} and the Bjorken 
sum~\cite{Bj EG1b}. These fits point out 
the importance of including a $Q^6$ 
term for $Q^2<0.1$ GeV$^2$. The $\chi PT$ calculations seems to agree best with the 
measurement of the Bjorken sum, in accordance with the discussion in~\cite{Burkert Delta}. 
Phenomenological models~\cite{AO},\cite{soffer} are in good agreement with the data
over the whole $Q^2$ range.

\section{The strong coupling at large distances}
A primary goal of the JLab experiments
was to map precisely the intermediate $Q^2$ range in order to shed light on the 
transition from short distances (where the degrees of freedom pertinent to the strong force
are the partonic ones) to large distances where the hadronic degrees of freedom are 
relevant to the strong force. One feature seen on Fig.~\ref{gammas} is that the transition
from small to large distances is smooth, e.g. without sign of a phase transition. This
fact can be used to extrapolate the definition of the strong force effective coupling
to large distances. Before discussing this, we first review the QCD coupling 
and the issues with calculating it at large distances.

In QCD, the magnitude of the strong force is given by 
the running coupling constant $\alpha_{s}$. 
At large $Q^2$, in the pQCD domain, $\alpha_{s}$ is well defined and 
is given by the series:
\vspace{-0.2cm}
\begin{eqnarray}
\mu\frac{\partial\alpha_{s}}{\partial\mu} =2\beta(\alpha_{s})
=-\frac{\beta_{0}}{2\pi}\alpha_{s}^{2}-\frac{\beta_{1}}{4\pi^{2}}
\alpha_{s}^{3}-\frac{\beta_{2}}{64\pi^{3}}\alpha_{s}^{4}-...
\label{eq:alpha_s beta serie}
\end{eqnarray}
Where $\mu$ is the energy scale, to be identified to $Q$. The 
first terms of the $\beta$ series are: 
$\beta_{0}=11-\frac{2}{3}n$ with $n$ the number of active quark 
flavors, $\beta_{1}=51-\frac{19}{3}n$ and $\beta_{2}=2857-
\frac{5033}{9}n+\frac{325}{27}n^{2}$.
The solution of the differential equation \ref{eq:alpha_s beta serie} is:
\small
\vspace{-0.2cm}
\begin{eqnarray}
\alpha_{s}(\mu)=\frac{4\pi}{\beta_{0}ln(\mu^{2}/\Lambda_{QCD}^{2})}
\times \label{eq:alpha_s} [1-\frac{2\beta_{1}}{\beta_{0}^{2}}
\frac{ln\left[ln(\mu^{2}/\Lambda_{QCD}^{2})\right]}{ln(\mu^{2}/
\Lambda_{QCD}^{2})}+ \nonumber \\
\frac{4\beta_{1}^{2}}{\beta_{0}^{4}ln^{2}
(\mu^{2}/\Lambda_{QCD}^{2})} \times \nonumber \\ 
\left(\left(ln\left[ln(\mu^{2}/
\Lambda_{QCD}^{2})\right]-\frac{1}{2}\right)^{2}+\frac{
\beta_{2}\beta_{0}}{8\beta_{1}^{2}}-\frac{5}{4}\right)]
\end{eqnarray}
\normalsize

\noindent Equation \ref{eq:alpha_s} allows us to evolve the different experimental 
determinations of $\alpha_{s}$ to a conventional scale, typically 
$M_{z_{0}}^{2}$. 
The agreement between the $\alpha_{s}$ obtained from different observables
demonstrates its universality and the validity of 
Eq. \ref{eq:alpha_s beta serie}. One can obtain  
$\alpha_{s}(M_{z_{0}}^{2})$ with doubly polarized DIS data 
and assuming the validity of the Bjorken sum. Solving 
Eq. \ref{eq:bj(Q2)} using the experimental value of 
$\Gamma_{1}^{p-n}$, and then using Eq. \ref{eq:alpha_s} provides 
$\alpha_{s}(M_{z_{0}}^{2})$.

Equation \ref{eq:alpha_s} leads to an infinite coupling at large distances, 
when $Q^2$ approaches $\Lambda^{2}_{QCD}$. 
This is not a conceptual problem since we are out of the validity 
domain of pQCD on which Eq. \ref{eq:alpha_s} is based.
But since data show no sign of discontinuity or phase transition 
when crossing the intermediate $Q^{2}$ domain, one 
should be able to define an effective coupling $\alpha_{s}^{eff}$ at 
any $Q^2$ that matches $\alpha_{s}$ at large $Q^{2}$ but stays finite at 
small $Q^{2}$. 

The Bjorken Sum Rule can be used to define $\alpha_{s}^{eff}$ at 
low Q$^{2}$. Defining $\alpha_{s}^{eff}$
from a pQCD equation truncated to first order (in our case Eq. (\ref{eq:bj(Q2)}: 
$\Gamma_{1}^{p-n}\equiv\frac{1}{6}(1-\alpha_{s,g_{1}}/\pi)$),
offers advantages. In particular, $\alpha_{s}^{eff}$ does not 
diverge near $\Lambda_{QCD}$ and is renormalization scheme independent. 
However, $\alpha_{s}^{eff}$ becomes dependent 
on the choice of observable employed to define it. If $\Gamma_{1}^{p-n}$ is 
used as the defining observable, the effective coupling is noted 
$\alpha_{s,g_{1}}$.  
Relations, called \emph{commensurate scale relations} \cite{Brodsky CSR}, 
link the different effective couplings so in principle one
effective coupling is enough to describe the strong force and the theory 
retains its predictive power. These relations are defined for short distances
and whether they extrapolate to large distances remains to be investigated. 

The choice of defining the effective charge with the Bjorken sum has many advantages: 
low $Q^2$ data exist and near real photons data from JLab is being 
analyzed~\cite{gdh neutron,gdh proton}. Furthermore, sum rules constrain
$\alpha_{s,g_1}$ at both low and large $Q^2$, as will be discussed in
the next paragraph. Another advantage is that, as discussed for the low $Q^2$ 
domain, the simplification arising in $\Gamma _1^{p-n}$ makes a quantity 
well suited to be calculated at any $Q^2$~\cite{Burkert Delta}. These simplifications 
are manifest 
at large $Q^2$ when comparing the validities of the Bjorken and Ellis-Jaffe sum rules.
It also simplifies Lattice QCD calculations in the intermediate $Q^2$ domain. 
Finally, it may be argued that $\alpha_{s,g_1}$ might be more directly 
comparable to theoretical calculations than other effective couplings extracted from other
observables: part of the coherent response of the nucleon
is suppressed in the Bjorken sum, e.g. the $\Delta$ resonance, so the non-resonant
background, akin to the pQCD incoherent scattering process, contributes especially
importantly to the Bjorken sum. This argument is reinforced if global duality works, 
a credible proposal since the  $\Delta$ resonance is suppressed.

The effective coupling definition in terms of 
pQCD evolution equations truncated to first order was proposed by 
Grunberg \cite{Grunberg}. Grunberg's definition is meant for
short distances but one can always extrapolated this definition and see 
how the resulting coupling compares to calculation
of $\alpha_{s}$ at large distances. 
Using Grunberg's definition at large distances entails
including higher twists in $\alpha_{s,g_{1}}$ in addition to the higher terms of the
pQCD series. 

It is common to fold the dynamics due to forces (here the Higher Twists) into an effective 
parameter so that the particle can be treated as free. It is interesting to review quickly 
the characteristics of such effective parameters, e.g. in the field of quantum 
electronics. There, near the energy extrema of electrons moving in a crystal, the 
effects of external forces applied to the crystal are folded into effective masses and the 
electron motions can be described using the free Schrodinger equation. Then, the 
(effective) mass of an electron becomes a tensor $m^*_{ij}$ (that depends
on the electron energy) rather than a scalar since the crystal lattice is not isotropic 
and the total acceleration depends on the lattice forces. $m^*_{ij}$ depends on the material
and, near an energy maximum, $m^*_{ii}$ is negative. Holes also have effective masses of 
opposits signs as for electrons. Effective masses are useful to determine quantities of interest
of a material, such as the quantum state densities, the speed of electric signals. or the surface 
of isoenergy. This illustrates the relevance of effective parameters, but also that we should 
not be shocked if our effective couplings depends on reactions or are negative.

\begin{center}
\includegraphics[width=5cm]{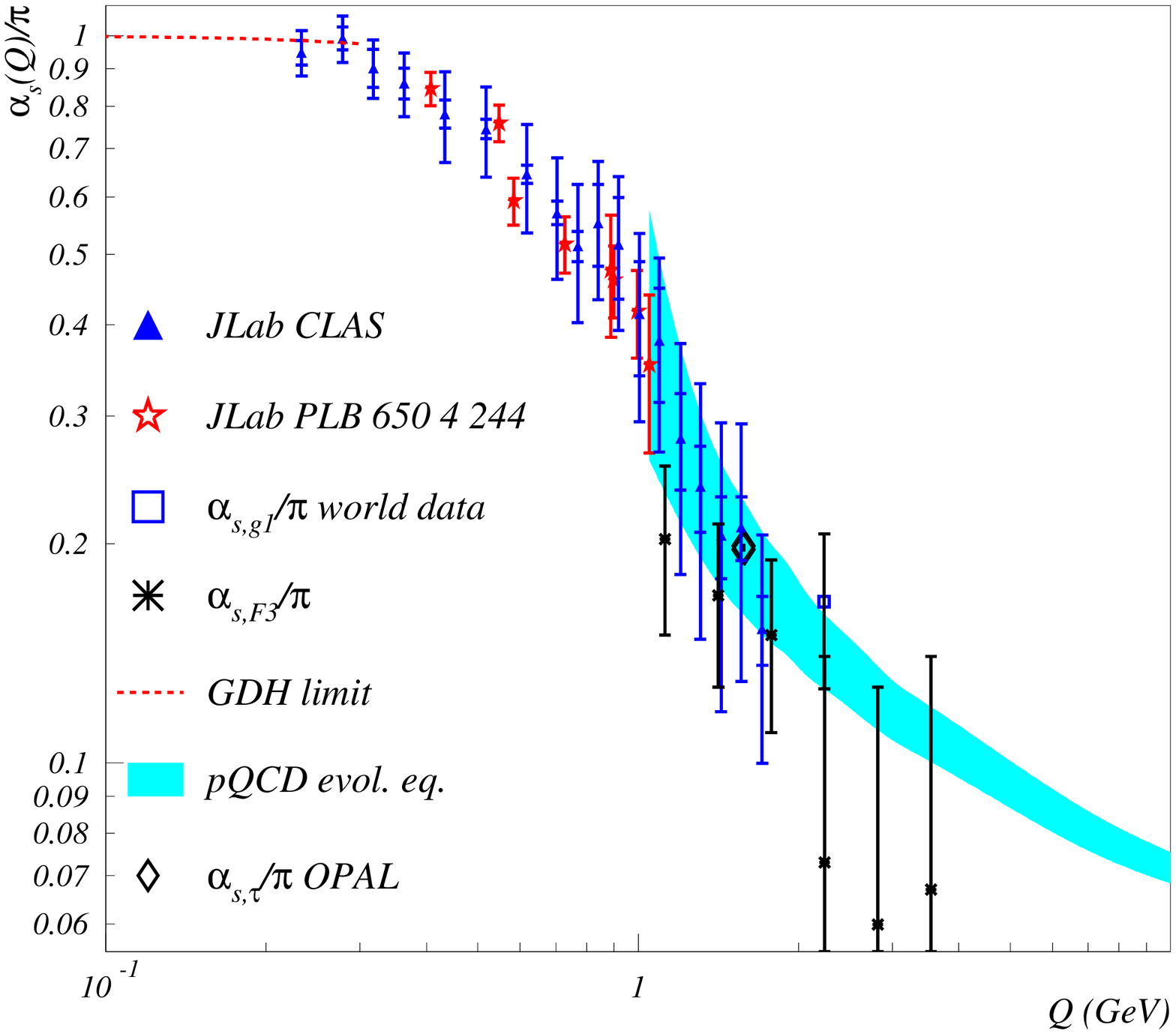}
\figcaption{\label{fig: alpha_s eff} 
Effective couplings extracted from different observables, see text for details.
The gray band indicates $\alpha_{s,g_{1}}$ extracted from the pQCD expression 
of the Bjorken sum at leading twist and third order in $\alpha_{s}$. The values of 
$\alpha_{s,g_{1}}/\pi$ extracted using the GDH sum rule is given by the red dashed line.}
\end{center}

Effective couplings have been extracted from different 
observables, see Fig. \ref{fig: alpha_s eff}. Values of $\alpha_{s,g_{1}}/\pi$ 
extracted from the world data on the Bjorken sum at 
$Q^{2}=5$ GeV$^{2}$ \cite{E155-E155x} and from JLab 
data~\cite{eg1a proton,Bj EG1b} have been compared using the commensurate scale relations~\cite{deur alpha_s^eff} to $\alpha_{s,\tau}$ extracted from the OPAL 
data on $\tau$ decay \cite{Brodsky CSR}, and $\alpha_{s,GLS}$ extracted 
using the Gross-Llewellyn Smith sum rule \cite{GLS} and its measurement by the 
CCFR collaboration \cite{CCFR}. There is good agreement between $\alpha_{s,g_{1}}$, 
$\alpha_{s,F_{3}}$ and $\alpha_{s,\tau}$. 

The GDH and Bjorken sum rules can be used to extract  
$\alpha_{s,g_{1}}$ at small and large $Q^{2}$ respectively 
\cite{deur alpha_s^eff}. This, together with the JLab data at 
intermediate $Q^{2}$, provides for the first time a coupling 
at any $Q^{2}$. A striking feature of Fig. \ref{fig: alpha_s eff} 
is that $\alpha_{s,g_{1}}$ becomes scale invariant at small $Q^{2}$. 
This was predicted by a number of calculations and it is 
known that color confinement leads to an
infrared fixed point \cite{irfp}, but it is the 
first time it is seen experimentally.
A fit of the $\alpha_{s,g_{1}}$ has been 
performed and is shown on Fig. \ref{fig: alpha_s eff4} 
(plain black line). 

\begin{center}
\includegraphics[width=6cm]{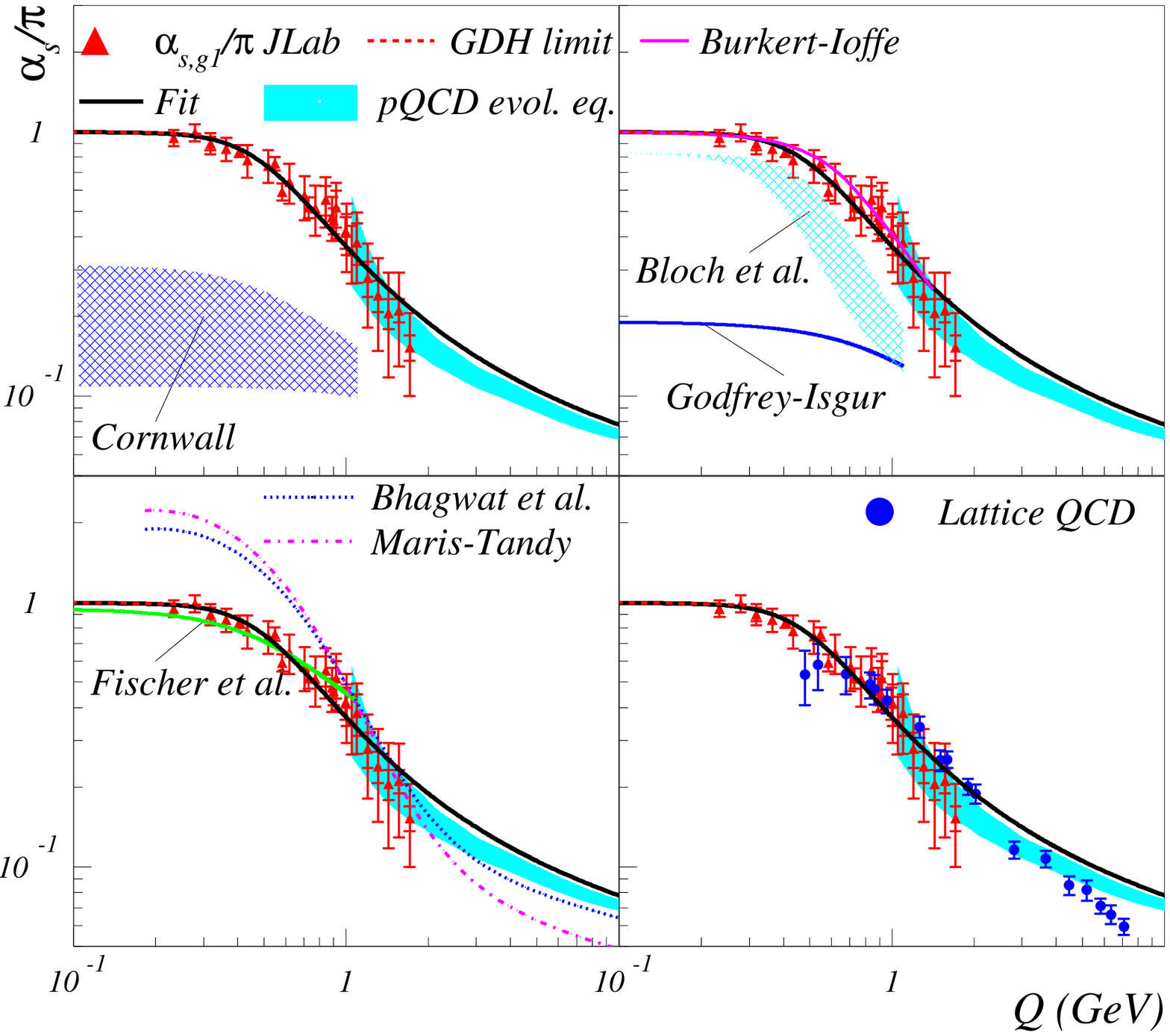}
\figcaption{\label{fig: alpha_s eff4} The effective coupling 
$\alpha_{s,g_{1}}$ compared to $\alpha_{s}$ calculations.}
\end{center}
There are several techniques used to predict 
$\alpha_{s}$ at small $Q^{2}$, e.g. lattice QCD, solving the 
Schwinger-Dyson equations, or choosing the coupling in a constituent 
quark model so that it reproduces hadron spectroscopy. However, the 
connection 
between  these $\alpha_{s}$ is unclear, in part because of the different 
approximations used. In addition, the precise relation 
between $\alpha_{s,g_{1}}$ (or any effective coupling defined using 
\cite{Grunberg} or \cite{Brodsky CSR}) and these computations is
unknown. Nevertheless, one can still compare them to see 
if they share common features. In Figure \ref{fig: alpha_s eff4}, $\alpha_{s,g_{1}}$ 
extracted from JLab data, its fit, and its 
extraction using the Burkert and Ioffe \cite{AO} model to 
obtain $\Gamma_{1}^{p-n}$ are compared to $\alpha_{s}$ calculations. 
The methods used are solving the Schwinger-Dyson equations 
(Top left: Cornwall \cite{Cornwall}; 
Top right: Bloch \cite{Bloch}; 
Bottom left: Maris-Tandy \cite{Tandy},Fischer, Alkofer, Reinhardt and Von 
Smekal \cite{Fischer}, and Bhagwat et al. \cite{Bhagwat}), 
$\alpha_{s}$ used in a quark constituent model 
(Godfrey-Isgure \cite{Godfrey-Isgur}) and Lattice QCD \cite{Furui} (bottom right). 
The calculations and $\alpha_{s,g_{1}}$ present 
a similar behavior. Some calculations, in
particular the lattice one, are in excellent agreement with 
$\alpha_{s,g_{1}}$. 

These works show that $\alpha_{s}$ is scale invariant (\emph{conformal 
behavior}) at small 
and large $Q^{2}$ (but not in the transition region between the 
fundamental description of QCD in terms of quarks and gluons 
degrees of freedom and its effective one in terms of baryons and mesons).
The scale invariance at large $Q^2$ is the well known asymptotic freedom. 
The conformal behavior at small $Q^{2}$ 
is essential to apply a property of \emph{conformal field theories} 
(CFT) to the study of hadrons: the \emph{Anti-de-Sitter 
space/Conformal Field Theory (AdS/CFT) correspondence} of Maldacena 
\cite{Maldacena}, that links a strongly coupled gauge field to weakly 
coupled superstrings states. Perturbative calculations are feasible in 
the weak coupling AdS theory. They are then projected on the AdS 
boundary, where they correspond to the calculations that would have
been obtained with the strongly coupled CFT. This opens the possibility 
of analytic non-perturbative QCD calculations \cite{ads/CFT}.

\section{Summary and perspectives}
We discussed the JLab data on moments of spin structure functions, in particular at large 
distances where we compared them to $\chi PT$, the strong force effective theory at large 
distances. 
The smoothness of $Q^2$-dependence of the moments when transiting from perturbative to 
the non-perturbative domain allows to extrapolate the definitions of effective strong 
couplings from short to large distances. Thanks to the data on nucleon spin structure 
and to spin sum rules, the effective strong coupling $\alpha_{s,g_1}$can be extracted 
in any regime of QCD. The question of comparing it with theoretical calculations of
$\alpha_{s}$ at low $Q^2$ is open, but such comparison exposes a
similarity between these couplings. Apart for the parton-hadron transition 
region, the coupling shows
that QCD is approximately a conformal theory. This is a necessary
ingredient to the application of the AdS/CFT correspondence that may 
make analytical calculations possible in the non-perturbative domain of QCD.

\end{multicols}

\clearpage

\end{document}